# Toward a Comprehensive Model of Snow Crystal Growth: 9. Characterizing Structure-Dependent Attachment Kinetics near -4 C

Kenneth G. Libbrecht

Department of Physics
California Institute of Technology
Pasadena, California 91125
kgl@caltech.edu

**Abstract.** In this paper I examine snow crystal growth near -4 C in comparison with a comprehensive model that includes Structure-Dependent Attachment Kinetics (SDAK). Together with the previous paper in this series that investigated growth near -14 C, I show that a substantial body of experimental data now supports the existence of pronounced "SDAK dips" on basal surfaces near -4 C and on prism surfaces near -14 C. In both cases, the model suggests that edge-associated surface diffusion greatly reduces the nucleation barrier on narrow facet surfaces relative to that found on broad facets. The remarkable quantitative similarities in the growth behaviors near -4 C and -14 C suggest that these two SDAK features arise from essentially the same physical mechanism occurring at different temperatures on the two principal facets. When applied to atmospheric snow crystal formation, this comprehensive model can explain the recurrent morphological transitions between platelike and columnar growth seen in the Nakaya diagram.

## ❄ INTRODUCTION

As suggested by its title, this series of papers represents my ongoing efforts to develop a comprehensive model of the physical dynamics underlying snow crystal growth. The research remains a work in progress, and some previous reviews of this subject can be found in [1954Nak, 1987Kob, 2005Lib, 2017Lib]. A substantive new book describing this subject in detail is currently in press [2021Lib].

As described in the above references, the Nakaya diagram has long presented an essential scientific puzzle in snow-crystal growth, as it illustrates a complex series of morphological transitions between thin platelike structures (appearing in air at temperatures near -14 C and again above -3 C) and slender columnar forms (near -4 C and again below -30 C). Discovered empirically in the 1930s by Ukichiro Nakaya and collaborators, it has proven quite challenge to develop a physical model that convincingly explains these remarkable changes in ice growth behaviors with temperature.

In [2019Lib1], I proposed a Comprehensive Attachment Kinetics (CAK) model describing a microscopic mechanism that can plausibly explain both the dominant morphological transitions in the Nakaya diagram and a larger body of later experimental data examining ice growth rates as a function of temperature and supersaturation over a broad range of conditions. The model borrows ideas from previous attempts to explain the Nakaya diagram [1958Hal, 1963Mas, 1982Kur, 1984Kur], most notably the hypothesis that the onset of surface premelting occurs at different temperatures on the basal and prism facets. But it additionally incorporates a key new hypothesis that edge-associated surface diffusion can effectively reduce the nucleation barrier on narrow faceted surfaces. This new mechanism, which I have been calling Structure Dependent Attachment Kinetics (SDAK) [2003Lib1], provides a robust connection between snow-crystal growth morphologies and the underlying molecular processes taking place at the ice surface.

Importantly, the CAK model makes a host of new predictions that are imminently testable using targeted experimental investigations, and the model seems to be holding up remarkably well to quantitative scrutiny so far [2019Lib2, 2020Lib, 2020Lib1]. The results presented here support the model further, revealing important physical similarities between the SDAK dips at -14 C and -4 C, as expected.

## ❄ The CAK Model near -4C

The Comprehensive Attachment Kinetics (CAK) model has been described in considerable detail in [2019Lib1, 2020Lib1, 2021Lib], so I will only provide a summary of its main features here, focusing on temperatures near -4 C. On broad faceted surfaces of effectively infinite extent (meaning that all edge effects are negligible), the attachment coefficients are given by

$$\alpha_x(\sigma_{surf}) = A_x e^{-\sigma_{0,x}/\sigma_{surf}} \quad (1)$$

where $x$ stands for either $basal$ or $prism$, $\sigma_{surf}$ is the water-vapor supersaturation near the ice surface, the nucleation parameter $\sigma_{0,x}$ derives from the terrace step energy at the ice/vapor interface, and $A_x$ depends on admolecule surface diffusion and other factors. This functional form is provided by terrace-nucleation theory [e.g., 1994Ven, 1996Sai, 1999Pim, 2002Mut], and the CAK model values of $\alpha_{basal}$, $\alpha_{prism}$, $A_{basal}$, and $A_{prism}$ as a function of temperature $T$ are shown in Figure 1.

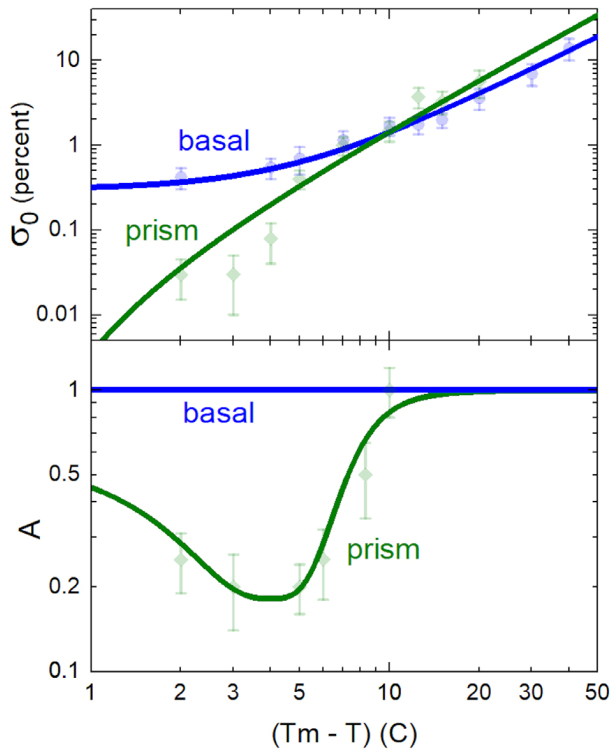

Figure 1: Model parameters for the growth of broad basal and prism facets, according to the CAK model. The nucleation parameter $\sigma_0$ derives from the terrace step energy at the ice/vapor interface, as dictated by classical nucleation theory. The $A$ parameter depends on admolecule surface transport and other factors that are difficult to determine accurately. These parameters were all estimated from measurements of ice growth as a function of temperature and supersaturation, as described in [2021Lib].

There is no theory in condensed-matter physics that provides these parameters from first principles, so the four curves in Figure 1 were all estimated from smoothed experimental data, as is discussed in detail in [2021Lib]. The asymptotic value of $A = 1$ indicates fast kinetics above the nucleation barrier, meaning $\alpha \to 1$ when $\sigma_{surf} \gg \sigma_0$.

Most of the experimental data supporting Figure 1 come from ice-growth measurements in near-vacuum conditions, but the available evidence also suggests that the presence of a background gas of air does not change any of these parameters appreciably [2021Lib]. Of course, because these model parameters were all derived from measurements, the curves in Figure 1 come with some degree of uncertainty, and not every corner of parameter space has been thoroughly examined by existing experiments. For the present discussion, however, these known uncertainties should not substantially affect our final conclusions. For this reason, we will assume that the parameter curves in Figure 1 are essentially perfect, as our focus in this paper is primarily on the basal SDAK effect near -4 C.

## The SDAK-1 Effect

A fundamental tenet of the CAK model is that the attachment kinetics on broad faceted surfaces cannot alone be sufficient to explain many aspects of snow crystal growth. Instead, the CAK model postulates that the attachment kinetics can be dramatically altered on narrow faceted surfaces, driven by the edge-related surface diffusion effects described in [2019Lib1, 2021Lib]. Specifically, surface transport from basal/prism corners can increase the admolecule density on nearby faceted surfaces (relative to the "normal" admolecule density on broad faceted surfaces), and thus effectively reduce the nucleation barriers on these surfaces. I refer to this phenomenon as Structure-Dependent Attachment Kinetics (SDAK) because the *mesoscopic* structure of the overall crystal morphology, specifically the location of facet edges, can substantially alter the attachment kinetics relative to what is found on broad faceted surfaces [2003Lib1].

Importantly, the CAK model stipulates that this SDAK effect appears at different temperatures on the basal and prism facets, driven by temperature-dependent surface premelting on these different surfaces. As described in [2019Lib1, 2021Lib], this first SDAK phenomenon replaces the $\sigma_0$ panel in Figure 1 with the small-facet curves shown in Figure 2.

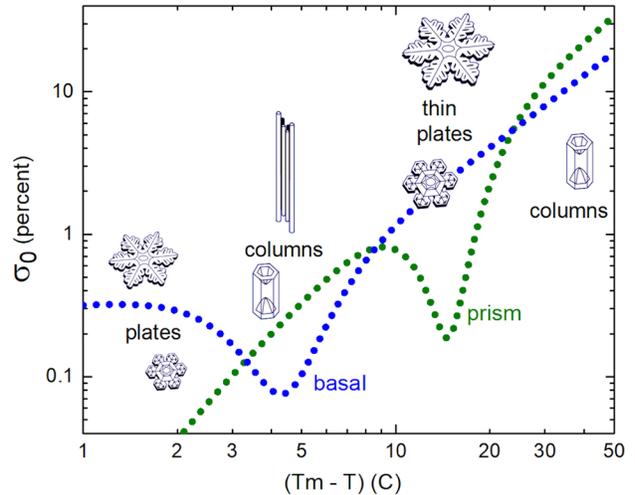

Figure 2: The CAK model effective nucleation parameters on narrow facet surfaces growing in air, including the "SDAK dip" in $\sigma_{0,prism}(T)$ near -14 C and the corresponding dip in $\sigma_{0,basal}(T)$ near -4 C. As illustrated here, the SDAK phenomenon on the principal facets can explain the transitions between platelike and columnar growth seen in the Nakaya diagram [2021Lib].

Unpacking the information contained in the curves in Figures 1 and 2, one can see that the CAK model contains just a few central ideas:

1) The attachment kinetics on broad faceted surfaces remains that shown in Figure 1. The parameters $\sigma_{0,basal}(T)$ and $\sigma_{0,prism}(T)$ derive from the (1D) terrace step energies, which are fundamental equilibrium properties of ice, much like the (2D) surface energies and (3D) latent heat. The step energies cannot yet be calculated accurately from first principles, but they can have been measured in experiments.

2) On narrow prism facets, the $\sigma_{0,prism}$ parameter exhibits a narrow "SDAK dip" near -14 C, which was quantified in [2020Lib1]. This dip does not result from a change in the step energy, but rather from edge-related admolecule transport that increases the nucleation rate above that given by terrace nucleation theory (as this theory strictly only applies when edge effects are negligible).
3) On narrow basal facets, the $\sigma_{0,basal}$ parameter exhibits a corresponding SDAK dip near -4 C, which we examine below. The CAK model proposes that both these dips are related to the onset of surface premelting, which occurs at different temperatures on the basal and prism facets.

Our understanding of the molecular dynamics of the ice surface is not sufficient to realistically quantify the SDAK-1 effect on the principal facets, or even to roughly estimate the temperatures of the two SDAK dips. Molecular dynamics simulations hold promise in this regard [2019Ben, 2020Llo], but testing the CAK model using *ab initio* calculations appears to be a somewhat distant goal. It is possible, however, to investigate the SDAK phenomenon using targeted experimental investigations, as was demonstrated near -14 C in [2020Lib1] and is further demonstrated near -4 C below.

When examining the CAK model, I often simplify the discussion by assuming that the SDAK-1 effect can be fully characterized by replacing the parameter $\sigma_{0,x}$ on broad facets with a new parameter $\sigma_{0,x,SDAK}$ on narrow facets, as illustrated in Figure 2. This is clearly something of an oversimplification, although theoretical considerations do suggest that an effective reduction of the nucleation barrier on narrow facets is a primary result of the SDAK phenomenon [2019Lib1]. Moreover, I have found that the functional form in Equation 1 can be used to characterize the experimental measurements reasonably well. Thus, from a purely empirical perspective, it makes some sense to define and discuss a single-valued $\sigma_{0,x,SDAK}(T)$ parameter, at least as a first step toward understanding the narrow faceted structures.

One reason that a discussion of a single-valued $\sigma_{0,x,SDAK}(T)$ function is fruitful stems from the morphological self-assembly that takes place when snow crystals grow in air. For example, dendritic structures appear at high supersaturations, and these typically assume a roughly parabolic tip structure because of diffusion effects, as dictated by the Ivantsov solution to the diffusion equation [2021Lib]. These structures tend to produce uppermost basal and prism terraces with widths that depend only on temperature and supersaturation, but not on the initial conditions of the crystal growth. As a result, the facet widths are not independent variables in the equations but assume certain fixed values that arise from the structural self-assembly present in diffusion-limited growth.

Put another way, a full theory of the SDAK phenomenon would provide a reduction of the nucleation barrier that depends on $T$, $\sigma_{surf}$, and the mesoscopic structure of the crystal, especially the width of the uppermost terrace surfaces. For reasons that are only partly understood at present [2019Lib1], it seems to work reasonably well to approximate the resulting growth behavior by replacing $\sigma_{0,x}$ by $\sigma_{0,x,SDAK}$ whenever the surface in question does not exhibit a broadly faceted structure. Although this must be an incomplete formulation of the problem, it gives us a useful starting platform for discussing experimental data.

### THE SDAK-2 EFFECT
At temperatures below -10 C, precision ice-growth measurements indicate that $A_{basal} \approx A_{prism} \approx 1$, and this allows for a relatively simple discussion of the SDAK-1 effect near -14 C [2020Lib1]. The situation is more complicated near -4 C, however, because $A_{prism}$ is substantially less than unity in that temperature range, as seen in Figure 1. Moreover, numerous experiments suggest that $A_{prism} \rightarrow 1$ on narrow prism facets at high $\sigma_{surf}$ in this temperature range, which we attribute to a second SDAK effect, which I call SDAK-2 [2019Lib2, 2020Lib].

The SDAK-2 phenomenon is not well understood at present, but there appears to be a strong correlation between $A_{prism}$ and facet width, suggesting that the mesoscopic crystal morphology again plays a role in defining the attachment kinetics contained in the $A_{prism}$ parameter, just as it does with SDAK-1. Observations suggest that $A_{prism} < 1$ on broad facet surfaces while $A_{prism} \to 1$ on fast-growing dendritic structures in air with narrow prism facets [2021Lib]. Specific examples at -5 C [2019Lib2] and at -2 C [2020Lib] support these assertions with quantitative data. Unfortunately, the facet width is inexorably linked with $\sigma_{surf}$ in these experiments, so it becomes difficult to disentangle the effects of facet width and surface supersaturation on growth rates. The SDAK-2 phenomenon is thus rather poorly understood, but it is also a relatively minor part of the CAK model.

It is useful to examine some specific cases of $\alpha_x(\sigma_{surf})$ at fixed temperature, and Figure 3 shows two examples that are relevant to the present paper. At -2 C, the attachment kinetics are mainly defined by the broad-facet curves, yielding simple platelike growth at low $\sigma_{surf}$. When dendritic structures emerge at higher $\sigma_{surf}$, the SDAK-2 effect increases $\alpha_{prism}$ on the uppermost prism terraces, so both the basal and prism kinetics exhibit $\alpha_x \to 1$ at the highest growth rates. A detailed comparison of the CAK model with a broad range of experimental data at -2 C is given in [2020Lib].

The CAK model at -5 C is more complicated than at -2 C, as both the SDAK-1 and SDAK-2 effects are present, as seen in Figure 3. While the basal SDAK-1 effect is essentially absent at -2 C, it is quite strong at -5 C, yielding the growth of columnar crystals at the latter temperature. It is possible, however, when $\sigma_{surf} \approx 0.1$ percent, to observe the growth of both platelike and columnar crystals simultaneously (brought about by different initial conditions), as was reported by Knight [2012Kni]. A detailed comparison of the CAK model with a broad range of experimental data at -5 C is given in [2019Lib2].

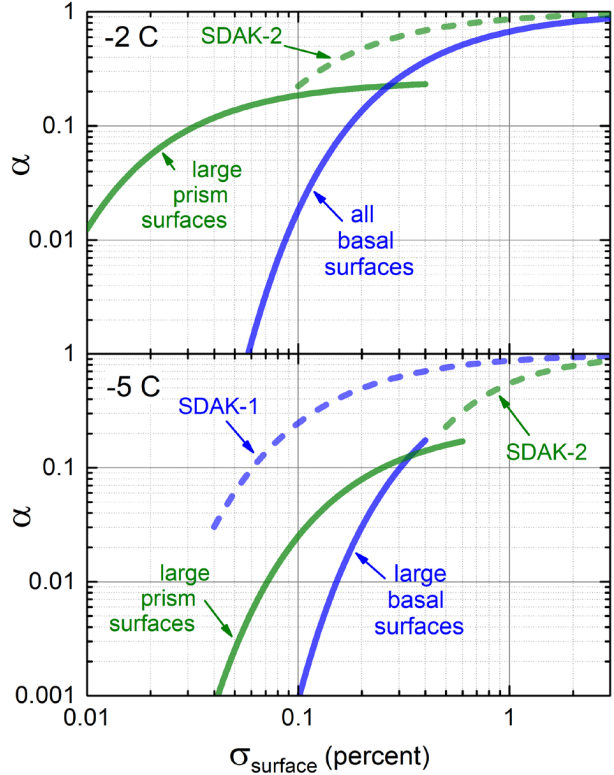

Figure 3: The CAK model at -2 C (top) at -5 C (bottom), showing $\alpha_x(\sigma_{surf})$ in both cases. The solid curves indicate broad-facet attachment kinetics, while the dashed curves represent the different SDAK-1 and SDAK-2 branches, as labeled. Note that the SDAK-1 effect applies to basal surfaces at -5 C, while the SDAK-2 effect always applies to prism surfaces.

In each of the graphs in Figure 3, the dotted curves are approximate single-valued functions of $\alpha_x(\sigma_{surf})$ representing the different SDAK effects. As described above, these specific "branches" in $\alpha(\sigma_{surf})$ space arise from self-assembly effects, so they apply mainly to snow crystal growth in air. At different air pressures, for example, we would expect these branches to shift somewhat. A full 3D computational model of diffusion-limited growth, incorporating the CAK model in the surface boundary conditions, would be needed to describe the full range of possible growth behaviors. This comprehensive numerical challenge is beyond the scope of the present paper, but it may be realized in the near future, given recent advances in computational techniques applied to diffusion-limited snow crystal growth [2009Gra, 2012Bar, 2017Dem, 2017Dem1].

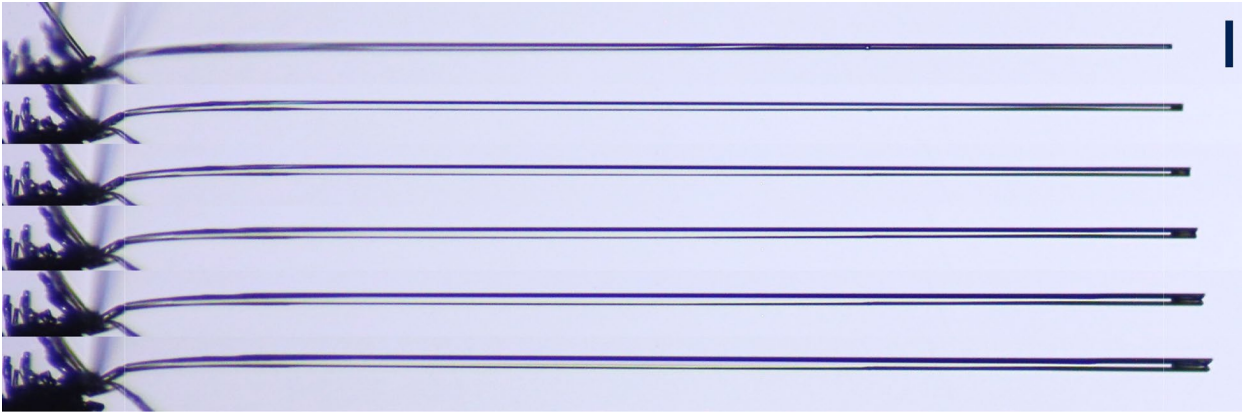

## Hypothesis Testing

At this juncture, it is useful to note that my primary goal in this paper is one of hypothesis testing, with the entire CAK model being the hypothesis in question. The CAK model makes numerous quantitative predictions regarding ice-growth behaviors that are best examined using targeted experimental investigations. To this end, I have performed a series of experimental measurements that focus on specific phenomena in the CAK model, and I analyze the measurements within the model framework. If the measurements generally agree with the model, reinforcing a self-consistent physical picture of the underlying ice-growth dynamics, then one can conclude that the experiments support the model.

On the other hand, if the model and the measurements conflict in any serious way, or yield physically unreasonable results, then one could conclude that the model has problems, perhaps to the point that the model would have to be rejected. Our overarching strategy with these experiments, therefore, is to assume that our hypothesis (the CAK model) is correct, and then see if this assumption seriously conflicts with experimental observations.

Figure 4: This composite image shows an initial ice needle (top) growing at -4 C with $\sigma_\infty \approx 8$ percent, over a period of 150 seconds. Near the end of the series, the columnar radius was 14 μm, the radial growth velocity was 49 nm/sec, the axial velocity was 490 nm/sec, and the columnar tip exhibited significant basal hollowing. The scale bar on the upper right is 100 μm long, and the axial growth was measured relative to the structure at the base of the needle, which was nearly constant with time.

## ❄ Growth at $\sigma_\infty$ = 8%

Our first experimental goal in this paper is to produce a quantitative estimate of $\sigma_{0,basal,SDAK}(T)$ as a function of temperature near -4 C, quantifying the estimated SDAK dip in the basal curve shown in Figure 2. The corresponding prism SDAK dip near -14 C was similarly investigated in [2020Lib1], so I used a similar experimental and analysis strategy near -4 C.

The presence of the SDAK-2 effect near -4 C complicates matters compared to -14 C, but this problem can be avoided with a judicious choice of experimental parameters. If $\sigma_{surf}$ remains sufficiently low, then the prism facets remain quite broad and the SDAK-2 effect is essentially absent. From our previous observations, we know that the SDAK-2 effect is relatively unimportant if $\sigma_{surf} \lesssim 0.1$ percent at -2 C [2020Lib] and $\sigma_{surf} \lesssim 0.3$ at -5 C [2019Lib2], as illustrated in Figure 3. Moreover, the SDAK-2 effect is associated with narrow prism structures, so the observation of broad prism facets is itself an indicator that the SDAK-2 effect is likely not present.

After a bit of trial-and-error, I found that the SDAK-2 effect could be avoided by observing the growth of ice needles in air under conditions with $\sigma_\infty \approx 8$ percent, where $\sigma_\infty$ is the supersaturation far from the growing crystals. For the temperature range $-1C > T > -10C$, this consistently yielded broad faceted prism surfaces at the needle tips, so I did a series of measurements at constant $\sigma_\infty$ over this temperature range.

Figure 4 shows a growth-run example, beginning with an "electric" ice needle in a linear diffusion chamber [2014Lib1, 2021Lib]. Using a series of photographs of the growing needle, the needle tip radius $R$ and overall length $H$ were measured as a function of time. The corresponding radial and axial growth rates, $v_R$ and $v_H$, were then determined at a fixed time, after the needle growth had stabilized with clean prism facets, but before the needle radius exceeded $R \approx 20$ microns.

The surface supersaturation $\sigma_{surf}$ near the needle tip was determined using the "witness surface" method applied to the prism growth [2020Lib1, 2021Lib]. In the absence of the SDAK-2 effect, the CAK model defines $\alpha_{prism}(\sigma_{surf})$ on broad prism facets from the broad-facet parameters in Figure 1. Assuming this aspect of the model is correct, the value of $\sigma_{surf}$ is easily extracted from the radial velocity $v_R$. I further assume that this value of $\sigma_{surf}$ also applies to the nearby basal surfaces (numerical modeling confirms that this is a reasonably good assumption), and one can then extract $\alpha_{basal}(\sigma_{surf})$ from the axial velocity $v_H$.

When the temperature was within a few degrees of -4 C with $\sigma_\infty \approx 8$ percent, the tip structure exhibited considerable basal hollowing, as illustrated in Figure 4. Because the basal surfaces were quite narrow on these crystals, the basal growth could not be interpreted using the large-facet CAK model, instead being subject to the basal SDAK-1 effect. Outside this temperature range, however, basal hollowing was essentially absent in the images, revealing broad basal facets at the needle tips. In these cases, the witness-surface analysis yielded the expected broad-facet $\alpha_{basal}$, in agreement with the CAK model. In other words, when the basal facets were broad, the measured $v_H/v_R$ ratio was in good agreement with the broad-facet CAK model. When basal hollowing was present and the uppermost basal terraces were narrow, $v_H$ was found to be considerably higher than expected for broad facets, supporting the basal SDAK-1 effect in the CAK model.

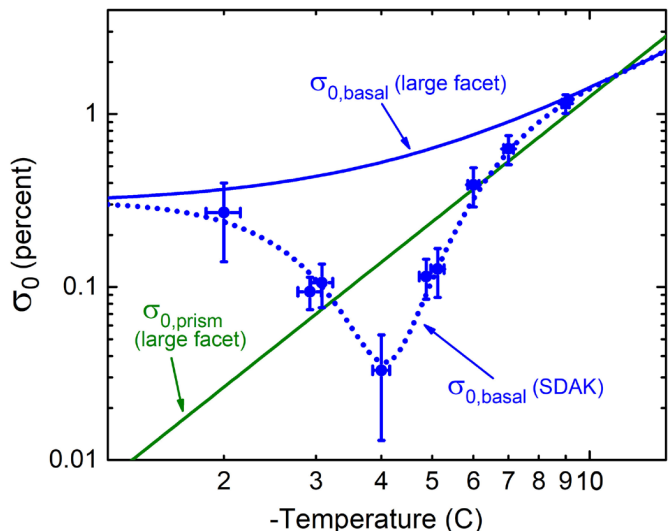

Figure 5: Measurements of the basal SDAK parameter $\sigma_{0,basal,SDAK}$ as a function of temperature, revealing a substantial "SDAK dip" centered near -4 C. For comparison, the solid lines show the CAK model parameters $\sigma_{0,basal}$ and $\sigma_{0,prism}$ for broad-facet growth.

As was done in [2020Lib1], I next extracted a value of $\sigma_{0,basal,SDAK}$ from each measurement of $\alpha_{basal}(\sigma_{surf})$, using the functional form

$$\alpha_{basal}(\sigma_{surf}) = e^{-\sigma_{0,basal,SDAK}/\sigma_{surf}} \qquad (2)$$

and Figure 5 shows a plot of $\sigma_{0,basal,SDAK}(T)$ together with $\sigma_{0,basal}(T)$ and $\sigma_{0,prism}(T)$.

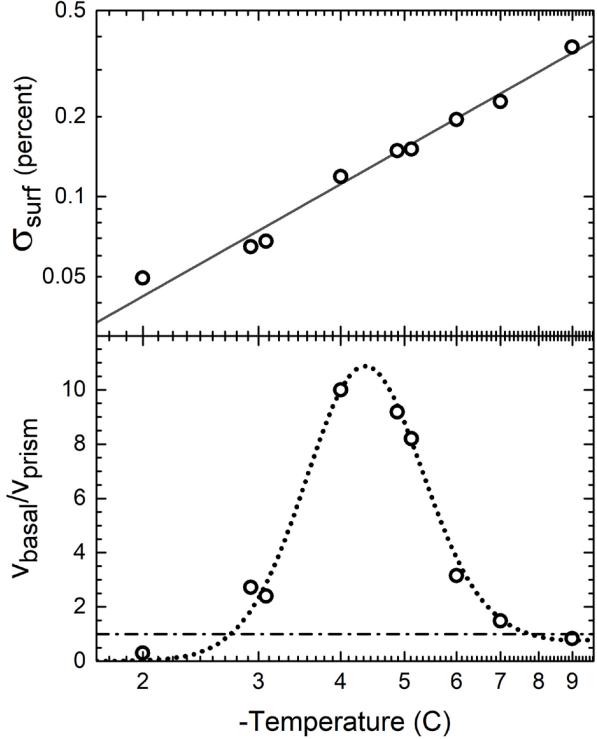

Figure 6: Additional data from growth with $\sigma_\infty = 8$ percent, showing the inferred $\sigma_{surf}$ from the witness-surface analysis (top) and the measured ratio of basal to prism growth velocities (bottom).

Figure 6 shows some additional data from this series of measurements, specifically the values of $\sigma_{surf}$ extracted from the witness-surface analysis and the measured (model-independent) velocity ratios. The change in $\sigma_{surf}$ with temperature was larger than expected in this experiment, apparently resulting from droplet nucleation in the diffusion chamber. While the chamber temperatures were set to give a constant $\sigma_\infty \approx$ 8 percent [2014Lib1, 2021Lib], this value is greater than $\sigma_{water}$ when $T > -8$ C. As the temperature increased, therefore, droplet nucleation became possible, and some slowly falling droplets were observed in the imaging system. These droplets likely lowered $\sigma_\infty$ from its calculated value, but the extent of this reduction is difficult to determine. Fortunately, the witness-surface analysis determines $\sigma_{surf}$ directly from the radial growth rate, so the actual value of $\sigma_\infty$ in the chamber is not critical in the analysis.

A first conclusion from this data set is that the measurements provide considerable support for the CAK model. The extent and placement of the basal SDAK dip confirm the model predictions, and the enhanced SDAK growth appears only on narrow basal surfaces, as expected. In contrast, when the basal hollowing is largely absent (far from -4 C in Figure 5), the growth becomes consistent with the broad-facet value $\sigma_{0,basal}$. If the CAK model were not correct, it is easy to imagine that the measurements would not have conformed so nicely to model expectations.

Note that it is not possible to fully isolate the different parts of the CAK model using these experimental measurements, because of the inherent difficulty in determining $\sigma_{surf}$ accurately around a growing crystal in air. The witness-surface analysis gets around this problem to a large degree, but this analysis strategy requires using part of the CAK model (the broad-facet value of $\alpha_{prism}(\sigma_{surf})$, in this case) to examine the lesser-known SDAK effect contained in $\sigma_{0,basal,SDAK}(T)$. The exercise is thus something of a "bootstrap" process, by which we can slowly build a general understanding of the CAK model.

Another conclusion is that these data provide a first experimental measurement of the basal SDAK dip near -4 C. The curves in Figure 2 were rough estimates based on morphological observations and related considerations, without quantitative confirmation. The curve in Figure 5, together with its analog measuring $\sigma_{0,prism,SDAK}$ near -14 C [2020Lib1], provide the first significant measurements of the SDAK-1 effect in the CAK model.

Finally, these measurements of $\sigma_{0,basal,SDAK}(T)$ bring us another step closer to creating realistic computational models of snow crystal growth, as such efforts require an accurate and quantifiable model of the attachment kinetics over a broad range of growth conditions. The CAK model may be

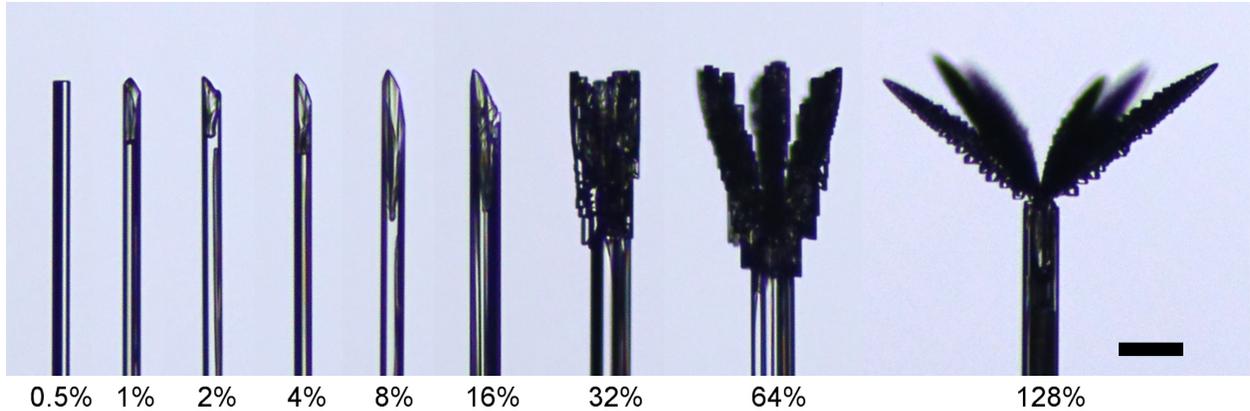

0.5%  1%  2%  4%  8%  16%  32%  64%  128%

Figure 7: This composite image shows a series of ice needles after growing at the $\sigma_\infty$ levels indicated. Note the progression from no basal hollowing (0.5%) to deep basal hollowing (8-16%) and then to complex dendritic structures. At 128%, the growth rates become consistent with $\alpha_{basal} \approx \alpha_{prism} \approx 1$.

sufficient to reproduce the broad variety of snow crystal growth morphologies observed, provided a set physically reasonable numerical algorithms can be devised. A suitable comprehensive numerical model of snow crystal growth has not yet been demonstrated but achieving this goal may be possible in the not-too-distant future, given the rapid development of available computational methods [2009Gra, 2012Bar, 2017Dem, 2017Dem1].

## ❋ Growth at -4 C

It is also instructive to examine growth at -4 C more closely, where the basal SDAK effect is most pronounced. To this end, I grew a series of ice needles as a function of supersaturation $\sigma_\infty$ at this temperature, analyzing the data using the same witness-surface method described above. Figure 7 illustrates how the growth morphology changes with increasing $\sigma_\infty$ at -4 C, progressing from solid columns to hollow columns to complex dendritic structures.

For each observed crystal, the measured axial and radial velocities at the outermost points represent the growth of the uppermost basal and prism facet surfaces. Even with a complex crystal morphology, these velocities must be determined by $\alpha_{basal}$ and $\alpha_{prism}$ on the uppermost facet surfaces. For broad facets, the CAK model predicts simple nucleation-limited growth, as described in Figure 1. For narrow surfaces, both the SDAK-1 and SDAK-2 effects come into play, and the simultaneous presence of both effects complicates the picture.

Figure 8 shows my interpretation of the velocity data, showing the inferred $\alpha_{basal}$ and $\alpha_{prism}$ values in different growth regimes. At low $\sigma_\infty$ in this graph, the columnar growth exhibited broad prism facets, so I assumed the CAK model with no SDAK effects, taking $\alpha_{prism} = 0.25 \exp(-0.14/\sigma_{surf})$ at -4 C, shown as the solid green curve in Figure 8. The measured $v_{prism}$ were then used to infer $\sigma_{surf}$, giving the dots shown on this curve. The measured $v_{basal}$ then gave $\alpha_{basal}$, giving the SDAK-1 basal points shown.

Because the SDAK-1 effect is so strong at -4 C, the columnar growth exhibited strong basal hollowing down to quite low supersaturations, as shown in Figure 7. Only at $\sigma_\infty = 0.5\%$, the lowest value measured, did the column exhibit what appeared to be a flat basal surface. Even then, it appears that the SDAK effect is not entirely absent, as the inferred $\alpha_{basal}$ in Figure 8 is substantially higher than that expected for broad basal facets. In contrast, platelike crystals were readily observed at -5 C in [2019Lib2], probably because the SDAK effect is not quite so potent as it is at -4 C.

For $\sigma_\infty$ in the range 1-16%, the growth morphology exhibited a "slanted" sheath-like hollow columnar form with a sharply tipped structure. A detailed analysis suggested that $\sigma_{surf}$ at the tip is somewhat higher than the value given by the witness-surface analysis, so this systematic effect was offset by introducing a factor of 1.3 increase for all slanted-needle morphologies. (A related witness-surface correction was described in [2020Lib].) This physically reasonable diffusion correction improved the general trend of the data, but it did not alter its most salient features. In particular, the rapid rise in $\alpha_{basal}$ with increasing $\sigma_{surf}$ seen in Figure 8 was especially insensitive to data analysis details. This rapid rise also agrees with common sense, as one expects $\alpha_{basal} \ll 1$ on faceted basal surfaces and $\alpha_{basal} \approx 1$ on sharp-tipped structures. Put another way, the rapid rise in $\alpha_{basal}$ seen in Figure 8 corresponds with the appearance of pronounced basal hollowing in the growth morphology.

Above $\sigma_{surf} \approx 0.2$ percent in Figure 8, the inferred $\alpha_{basal} \approx 1$ values correspond with sharply tipped dendritic morphologies, again confirming one's general intuition of fast kinetics on rounded structures. In contrast, the prism surfaces remained faceted up to $\sigma_\infty \approx 64\%$, suggesting $\alpha_{prism} < 1$. The SDAK-2 effect introduces a significant complication in the analysis, however, limiting what can be learned at high $\sigma_{surf}$. In Figure 8, I estimated the SDAK-2 curve, building on results at -2 C [2020Lib] and -5 C [2019Lib2], and used the usual witness-surface analysis to produce the inferred $\alpha_{basal}$ values shown. Alternatively, one could assume $\alpha_{basal} \approx 1$ and measure $\alpha_{prism}$ instead. Both approaches give essentially the same result, with the general trends shown in Figure 8.

At high supersaturations, the data all point to $\alpha_{basal} \approx \alpha_{prism} \approx 1$ at the highest $\sigma_{surf}$,

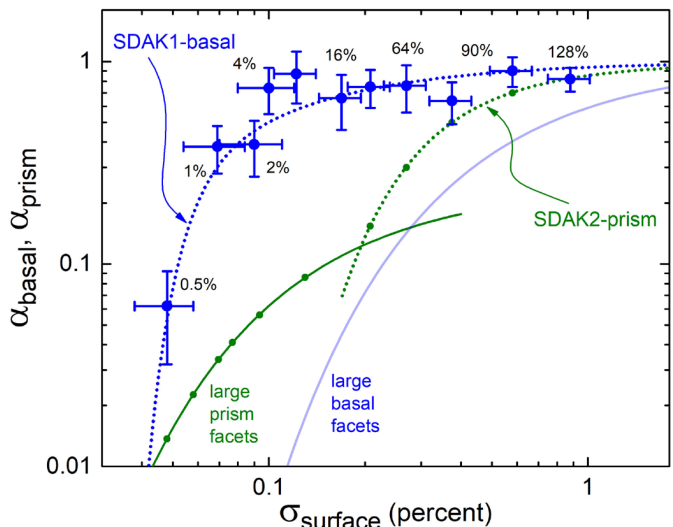

Figure 8: Inferred measurements of $\alpha_{basal}$ and $\alpha_{prism}$ as a function of $\sigma_{surf}$ at -4 C, as described in the text. Note the rapid rise in $\alpha_{basal}$ at $\sigma_{surf} \approx 0.5$ percent, signifying the onset of the SDAK effect, also causing the onset of basal hollowing at this same $\sigma_{surf}$. The simultaneous presence of the SDAK-1 and SDAK-2 effects complicates matters at high $\sigma_{surf}$, but the data seem to indicate $\alpha_{basal} \approx \alpha_{prism} \approx 1$ at the highest $\sigma_{surf}$.

where the growth morphology exhibits sharply tipped dendritic structures with $v_R \approx v_H$. This behavior appears over a large range of temperatures [2021Lib], consistent with one's expectation that $\alpha \to 1$ in the absence of any appreciable nucleation barrier.

## ❄ Conclusions

My primary conclusion in this paper is that the CAK model does a satisfactory job describing the ice/vapor attachment kinetics near -4 C. By assuming that the model is generally correct and analyzing the data accordingly, a physically sensible picture of snow crystal growth emerges, with no obvious inconsistencies or unphysical results. The overall trends in $\alpha_{basal}$ and $\alpha_{prism}$ generally agree with expectations from the model, with the predicted SDAK behaviors. Importantly, I found no obvious "showstoppers" that strongly disagree with the basic tenets of the CAK model.

One especially pleasing feature of the CAK model is that it quantifies the morphological transitions seen in the Nakaya diagram, presenting many opportunities for comparison with targeted experimental investigations. Moreover, as discussed above, changes in the inferred $\alpha_{basal}$ and $\alpha_{prism}$ with both temperature and supersaturation are observed to be accompanied by corresponding morphological changes, as predicted by the SDAK-1 and SDAK-2 effects in the CAK model.

Comparing the results presented here with the data near -14 C presented in [2020Lib1], it becomes difficult to avoid the conclusion that the basal SDAK dip near -4 C and the prism SDAK dip near -14 C are likely the same physical phenomenon, just occurring on different faceted surfaces. The general behaviors near these two temperatures are remarkably similar, including both the SDAK dip behaviors and the $\alpha_{basal,SDAK}(\sigma_{surf})$ and $\alpha_{prism,SDAK}(\sigma_{surf})$ curves. The CAK model predicts this, and the data clearly exhibit the expected similarities.

Considering the various experimental results together, we see that the CAK model provides a plausible and self-consistent physical picture of the processes underlying the long-puzzling Nakaya diagram. The broad-facet behavior alone is sufficient to explain the formation of plates at $T$>-3C and columns at $T$<-30C, as seen from the plot of $\sigma_{0,basal}$ and $\sigma_{0,prism}$ in Figure 1. The fact that $A_{prism} < 1$ at high temperatures complicates the picture, but only to a minor extent. Adding the basal SDAK dip yields strong columnar growth near -4 C, while the prism SDAK dip produces strong platelike growth near -14 C. Putting these pieces together thus yields the main temperature trends seen in the Nakaya diagram.